# Effects of a Prompt Engineering Intervention on Undergraduate Students' AI Self-Efficacy, AI Knowledge, and Prompt Engineering Ability: A Mixed Methods Study

David James Woo [a, *], Deliang Wang [b], Tim Yung[c], and Kai Guo[d]

[a] Precious Blood Secondary School, Hong Kong, China

[*] Corresponding author

- Postal address: Precious Blood Secondary School, 338 San Ha Street, Chai Wan, Hong Kong, China

- Email address: net_david@pbss.hk

- ORCID: https://orcid.org/0000-0003-4417-3686

[b] The University of Hong Kong, Hong Kong, China

[c] The University of Hong Kong, Hong Kong, China

[d] The University of Hong Kong, Hong Kong, China

**Authors' information**

*David James Woo* is a secondary school teacher. His research interests are in artificial intelligence, natural language processing, digital literacy, and educational technology innovations. Email address: net_david@pbss.hk ORCID: https://orcid.org/0000-0003-4417-3686

*Deliang Wang* is a Ph.D. student in the Faculty of Education at The University of Hong Kong. His research directions include artificial intelligence in education, explainable artificial intelligence, and educational data mining. His recent publications have appeared in international peer-reviewed journals and conferences, such as *Education and Information Technologies*, *International Journal of Educational Research*, *International Journal of Artificial Intelligence in Education*, *International Conference on Artificial Intelligence in education*, and *International Conference on Educational Data Mining*. Email address: wdeliang@connect.hku.hk ORCID: https://orcid.org/0009-0008-6488-0234

*Tim Yung* is a departmental lecturer and coordinator of the MA in Hong Kong History at the Department of History at the University of Hong Kong. While teaching a range of topics, his research interests include Hong Kong history, Chinese Christianity, empire and decolonization, and academic discourse socialization. His research has been published in *Studies in Church History*, among others. Email address: timyung@connect.hku.hk ORCID: https://orcid.org/0000-0001-5160-4232

*Kai Guo* is a Ph.D. candidate in the Faculty of Education at the University of Hong Kong. His research interests include technology-enhanced language learning, computer-supported collaborative learning, artificial intelligence in education, and gamification in education. His recent publications have appeared in international peer-reviewed journals such as *Computers & Education*, *Education and Information Technologies*, *Interactive Learning Environments*, *Journal of Educational Computing Research*, *TESOL Quarterly*, and *Assessing Writing*. Email address: kaiguo@connect.hku.hk ORCID: https://orcid.org/0000-0001-9699-7527

**Abstract**

Prompt engineering is critical for effective interaction with large language models (LLMs) such as ChatGPT. However, efforts to teach this skill to students have been limited. This study designed and implemented a prompt engineering intervention, examining its influence on undergraduate students' AI self-efficacy, AI knowledge, and proficiency in creating effective prompts. The intervention involved 27 students who participated in a 100-minute workshop conducted during their history course at a university in Hong Kong. During the workshop, students were introduced to prompt engineering strategies, which they applied to plan the course's final essay task. Multiple data sources were collected, including students' responses to pre- and post-workshop questionnaires, pre- and post-workshop prompt libraries, and written reflections. The study's findings revealed that students demonstrated a higher level of AI self-efficacy, an enhanced understanding of AI concepts, and improved prompt engineering skills because of the intervention. These findings have implications for AI literacy education, as they highlight the importance of prompt engineering training for specific higher education use cases. This is a significant shift from students haphazardly and intuitively learning to engineer prompts. Through prompt engineering education, educators can facilitate students' effective navigation and leverage of LLMs to support their coursework.

**Keywords**

AI self-efficacy; AI knowledge; prompt engineering; undergraduate students; AI education; ChatGPT

## 1. Introduction

The rapid development of artificial intelligence (AI) has had a profound impact on economies, employment, and various aspects of human society (UNESCO, 2019). In light of this, there is a growing need to enhance students' AI literacy to prepare them for an AI-powered society. According to Ng et al. (2021), AI literacy encompasses several important aspects, including understanding the fundamental functions of AI and its practical applications, applying AI knowledge and concepts in different scenarios, evaluating AI applications, and considering the accountability, transparency, ethics, and safety of AI applications.

The emergence of large language models (LLMs) like ChatGPT has significantly impacted higher education (Baig & Yadegaridehkordi, 2024). *Prompt engineering*, or creating effective instructions to interact with LLMs, has become a crucial skill that students must acquire to succeed in both academic and workplace settings. However, there is a limited understanding of how learning prompt engineering can impact students' AI literacy. Therefore, this study aims to examine the effects of a prompt engineering intervention on undergraduate students' AI self-efficacy, AI knowledge, and ability to craft prompts, in planning an essay with ChatGPT in a history class.

## 2. Literature Review

### 2.1. AI Self-Efficacy

Self-efficacy, as defined by Bandura (1991), is the belief in one's ability to achieve desired goals. It encompasses a sense of competence and confidence in successfully performing tasks. Individuals with high perceived self-efficacy are more likely to believe in their capability to accomplish given tasks, while those with lower self-efficacy may perceive tasks as more challenging. Additionally, technology self-efficacy is the belief in one's ability to interact with technological systems successfully (Kraus et al., 2021). In this way, we define AI self-efficacy as the belief in one's competence to interact with AI effectively. That competence encompasses confidence in using AI tools and making informed decisions when interacting with AI.

AI self-efficacy can influence various aspects related to students and AI. For instance, Asio and Gadia (2024) highlighted the significance of AI literacy and AI self-efficacy as predictors of student attitudes toward AI. Additionally, AI self-efficacy has been shown to impact students' acceptance of and intention to use AI technologies (Chai et al., 2021; Chen et al., 2024). In Kwak et al. (2022) involving 189 nursing students, higher AI self-efficacy scores were associated with stronger behavioral intentions to use AI-based healthcare technology, with third and fourth-year nursing students exhibiting higher self-efficacy scores compared to first and second-year students. Moreover, AI self-efficacy has been linked to students' learning achievements (Liang et al., 2023; S. Wang et al., 2023). Overall, these studies emphasize how AI self-efficacy shapes how students will interact with and benefit from AI use in schools and beyond.

Researchers have developed tools to measure students' AI self-efficacy. For instance, Morales-García et al. (2024) conducted an instrumental study with 469 medical students and adapted a six-item General Self-Efficacy Scale to measure self-efficacy in using AI among university students. The scale demonstrated a unidimensional structure with excellent fit indices. Additionally, the scale maintained its structure and meaning across genders, showing factorial invariance. Moreover, Y.-Y. Wang and Chuang (2024) developed a validated AI self-efficacy scale to measure individuals' perceived self-efficacy in using AI technologies. Their analysis of 314 responses revealed that AI self-efficacy consists of four factors: assistance, anthropomorphic interaction, comfort with AI, and technological skills. The scale, comprising 22 items, demonstrated good fit, reliability, convergent validity, discriminant validity, content validity, and criterion-related validity. The development and validation of these scales provide researchers and educators with effective tools for assessing students' beliefs in their own capabilities to use AI technologies.

Other studies have explored the factors that influence AI self-efficacy. Importantly, S. Wang et al. (2023) collected 561 questionnaire responses from students in Chinese higher education institutes that implemented AI technologies. Their study highlighted the significant influence of higher education institutes' AI capability, which encompassed resources (e.g., data, technical, and basic resources), skills (e.g., technical skills, teaching applications, and collaboration competencies), and consciousness (e.g., reform, and innovation consciousness), on their students' AI self-efficacy. As a result, higher education institutions have designed and implemented AI literacy courses. For example, Kong et al. (2021) designed, implemented and evaluated a 7-hour, AI literacy course in a flipped classroom mode. They assessed whether 120 students from various disciplines could develop a conceptual understanding of AI through the course. The pre-course and post-course survey results showed significant progress in participants' understanding of AI concepts, and students reported feeling empowered to work with AI. Thus, AI literacy courses can equip students with vital knowledge and skills to engage with AI technology effectively. As AI in society has rapidly changed not least through generative AI, LLMs and ChatGPT, higher education must revise their courses and incorporate new knowledge and skills for students' effective interaction with these emergent technologies.

## 2.2. AI Knowledge

AI knowledge refers to an individual's understanding of AI principles, applications, and technologies. It encompasses familiarity with AI concepts such as machine learning, neural networks, and natural language processing, as well as the ability to apply this knowledge to solve real-world problems and engage critically with AI-related issues. Studies have indicated that students' knowledge of AI varies significantly across different age groups and educational levels (Ng et al., 2023). Typically, younger students tend to possess a more basic understanding of AI, while older students and those exposed to AI-related coursework demonstrate a deeper comprehension of AI principles and applications. For instance, Su et al.'s (2023) recent review of AI education in K-12 contexts outlined a progression where kindergarten students focus on basic AI concepts (Williams et al., 2019), primary school students engage in hands-on activities involving machine learning (Toivonen et al., 2020), and secondary school students delve into co-designing and implementing machine learning applications (Vartiainen et al., 2020).

Effective pedagogical approaches for teaching AI to students have become a focal point of research interest. Strategies like inquiry-based learning (Ng et al., 2022), project-based learning (Toivonen et al., 2020), and learning by design (Shamir & Levin, 2022) have been identified as effective methods for enhancing students' understanding of AI concepts. However, assessing students' AI knowledge presents challenges because of AI's multidisciplinary nature and its rapid technological advancements. Ng et al.'s (2023) recent review on AI education in secondary schools highlighted various assessment techniques, with questionnaires being one of the most commonly used methods. For example, Pinski and Benlian (2023) developed a questionnaire to measure AI competence, assessing two aspects of AI knowledge: AI actor knowledge and AI steps knowledge. AI actor knowledge includes understanding of the roles of AI technology and human actors in human-AI collaboration and interaction, while AI steps knowledge encompasses AI input, AI processing, and AI output knowledge, reflecting students' understanding of AI processes and applications.

## 2.3. Prompt Engineering

Prompt engineering refers to crafting appropriate instructions for an LLM so that the LLM generates what the user wants (Liu et al., 2021). Since the quality of a prompt impacts the quality and relevance of an LLM's output (White et al., 2023), prompt engineering is an essential skill for effective interaction with ChatGPT and other LLM-based chatbots.

Computer science research has identified effective strategies for prompt engineering. These include prompting chatbots for a wide range of tasks for which the chatbots have been fine-tuned with human feedback (Ouyang et al., 2022), and following a catalog of strategies shown to improve chatbot output (White et al., 2023). Specific strategies include chain-of-thought prompting where the user either writes intermediate reasoning steps that guide the chatbot or asks the chatbot to explain its reasoning step-by-step before answering (Wei et al., 2023); and retrieval-augmented generation (RAG) where reference texts are given to the chatbot to improve its performance and to reduce hallucination (Lewis et al. 2021; Gao et al., 2024). Additionally, the in-context learning strategy refers to the user giving one, few or many examples to a chatbot to guide the chatbot's understanding of the desired output content and format (Agarwal et al., 2024; Fang et al., 2024). Besides, the producers and aggregators of LLM-based chatbots such as Anthropic (Prompt engineering, n.d.), OpenAI (OpenAI Platform, n.d.) and POE (Best practices for text generation prompts, n.d.) suggest prompt engineering strategies. These include chain prompting, where the user breaks down a complex task or text into smaller, more manageable chunks, and adding delimiters to help a chatbot to understand distinct parts of an input.

Although people may write effective prompts, doing so is not straightforward and easy for many (Zhou et al., 2023). Non-experts may learn prompt engineering intuitively when necessary but never robustly design and systematically test prompts (Zamfirescu-Pereira et al., 2023). In the context of higher education, although students increasingly use LLM-based chatbots, how they learn to engineer prompts is poorly understood. Researchers have explored students' intuitive behavior with LLM-based chatbots but not whether prompt engineering is a trainable skill for students (Knoth et al., 2024). Besides, educators have proposed prompt engineering frameworks for higher education (Eager & Brunton, 2023;

Lo, 2023) but that knowledge appears more grounded in those educators' tinkering than in computer science research.

Ultimately, if students adopt effective prompt engineering strategies, they may perform learning tasks better. Besides, that may enhance those students' AI self-efficacy. Since there is limited understanding of how teaching prompt engineering influences students' AI self-efficacy and their generative AI knowledge, especially about prompts, this study undertakes a structured intervention in prompt engineering to investigate its impact on undergraduate students' AI self-efficacy, generative AI knowledge, and their ability to engineer prompts. By bridging this gap, we can better incorporate prompt engineering effectively into AI literacy or content-specific courses, ensuring students gain the skills needed to navigate the evolving landscape of AI applications.

## 2.4. The Present Study
In this study, we implemented a prompt engineering workshop in a history course at a Hong Kong university. The workshop aimed to guide students in utilizing generative AI to assist with their academic writing tasks. This specific context enabled us to investigate the impact of prompt engineering education on students' AI self-efficacy, AI knowledge, and prompt engineering skills. The following research questions (RQs) were explored:

1) How does a prompt engineering workshop impact undergraduate students' AI self-efficacy?
2) How does the workshop influence students' knowledge of generative AI?
3) How does the workshop affect students' ability to engineer effective prompts for their academic writing tasks?

## 3. Material and Methods

### 3.1. Participants
The study's participants were an opportunistic sample of 27 undergraduate students enrolled in a credit-bearing history course at a Hong Kong university undertaken for three weeks in the summer term. The students ranged from first-year to final-year students and from 18 to 24 years old. Students were pursuing degrees in the arts, economics and finance, science, social sciences, and law. Two participants joined the course on a non-credit-bearing basis as visiting students. 17 of the 27 participants were completing the course to secure sufficient credits to graduate after summer. 23 of 27 participants expressed that their interest in Hong Kong history motivated them to enroll in the course. Experience in academic writing for arts and humanities varied depending on students' academic background. Only 2 of 27 had taken History courses during their undergraduate studies. Pseudonyms are used in this article.

### 3.2. Workshop Context and Design
The third author was the course instructor and wanted students to acquire generative AI knowledge and skills that could facilitate students' successful course completion and encourage their ethical use of AI. Thus, the third author contacted the first author to co-design a workshop titled, "How To Use ChatGPT To Plan A History Final Essay." For the workshop's learning design (see Table 1), the authors developed modules about process writing (Hyland, 2003), an introduction to generative AI including chatbot naming convention, and prompt engineering strategies, including in-context learning, chain prompting, chain-of-thought prompting and RAG. Lastly, the two authors designed a guided practice module where students would work with an instructor to apply prompt engineering strategies to plan the course's final essay task (see Appendix A). To support the guided practice section, the two authors prepared reference texts such as the course outline, final essay questions, model essays and academic sources from the course reading list that students could include in their prompts.

**Table 1**. Workshop learning design.

| Title | How to use ChatGPT to plan a history final essay |
|---|---|
| Time | 95 minutes |
| Purpose | To prompt AI to plan a history final essay given a) the task, b) the rubric and c) model essays |
| Intended learning outcomes (LT) | 1. To develop generic competence (i.e. knowledge, skills and attitude) to use ChatGPT<br>a. To apply different strategies for interacting with ChatGPT so ChatGPT delivers more valid and detailed essay plans.<br>b. To create a specialized chatbot that automates history essay planning. |
| Learning activities (minutes) | 1. Pre-workshop questionnaire<br>2. Pre-workshop prompt exercise (5 minutes)<br>3. Instructor introduction (5 minutes)<br>4. Process writing (5 minutes)<br>5. AI chatbot nomenclature: chatbots; versions; parameters; tokens; and context length etc. (10 minutes)<br>6. Prompt engineering: functions; context examples and strategies (15 minutes)<br>7. Guided practice (45 minutes)<br>i. Introduction to a history student's use case to have ChatGPT plan a history essay based on the task, the rubric and model essays. Divide cohort into novice group (ii) and non-novice group (v)<br>ii. Chain prompting<br>iii. Provide reference texts and use delimiters<br>iv. Systematically test<br>v. Use external tools<br>8. Post-workshop prompt exercise (5 minutes)<br>9. Post-workshop questionnaire (5 minutes) |
| Materials (written language) | Google Drive folder:<br>1. Pre-workshop questionnaire<br>2. Prompt library<br>3. Slide deck<br>4. Worksheets<br>5. Model essays with grades and XML tags<br>6. Final essay task<br>7. Course outline with task and rubric<br>8. Further reading<br>9. Post-workshop questionnaire |
| Instructional language | English |

## 3.3. Data Collection

The workshop was a one-time, 100-minute intervention held on July 3, 2024 during the course. At that time, students had attended the course for a week and a half, and had one and a half weeks remaining in the course. The first author was the workshop instructor and the third author also attended.

Table 2 chronologically sequences the data sources and their purposes. The study follows a convergent parallel design (Creswell & Clark, 2017): quantitative and qualitative data were collected simultaneously, analyzed separately and their results merged. Those quantitative and qualitative data were a pre-workshop questionnaire and a pre-workshop prompt library, respectively, in sequence 1, and a post-workshop questionnaire and post-workshop prompt library in sequence 2. Furthermore, the study follows an embedded design in that we followed quantitative data collection and analysis on the workshop's impact on AI self-efficacy and generative AI knowledge with qualitative data collection and analysis that can help explain the quantitative findings. That supplemental, qualitative data comprised end-of-term reflections in sequence 3. Thus, we collected different data types to deepen our understanding of the prompt engineering workshop by comparing results. For instance, the qualitative data could validate student self-reports of efficacy or provide an alternative perspective.

**Table 2**. Data sources and purposes.

| Sequence | Data Source | Purpose | Related RQs |
|---|---|---|---|
| 1 | Pre-workshop questionnaire | To establish a quantitative baseline for AI self-efficacy and generative AI knowledge | 1, 2 |
| | Pre-workshop prompt library | To establish a qualitative baseline for the ability to engineer effective prompts | 3 |
| 2 | Post-workshop questionnaire | To measure the workshop's impact on AI self-efficacy and generative AI knowledge | 1, 2 |
| | Post-workshop prompt library | To explore the workshop's impact on the ability to engineer effective prompts | 3 |
| 3 | End-of-term reflections | To provide further insights into AI self-efficacy and generative AI knowledge | 1, 2 |

### 3.3.1. Quantitative Data

Questionnaires were used to collect quantitative data on students' AI self-efficacy and knowledge of generative AI before and after the workshop. To assess students' AI self-efficacy, we adapted the scale developed and validated by Y.-Y. Wang and Chuang (2024), which measures individuals' self-efficacy in utilizing AI technologies. The AI self-efficacy questionnaire specifically evaluated students' *comfort levels with generative AI* and *technological skills*, focusing on their emotional awareness when interacting with generative AI and their confidence in using this technology. The questionnaire employed a seven-point Likert scale and included six items to measure comfort with AI (e.g., "*When interacting with ChatGPT and other POE chatbots, I feel very calm*") and four items to assess technological skills (e.g., "*When using ChatGPT and other POE chatbots, I am not worried that I might press the wrong button and cause risks*"). With a Cronbach's alpha value of 0.895, the instrument demonstrated high reliability in this study.

To assess students' knowledge of generative AI, we adapted the scale developed and validated by Pinski and Benlian (2023), which measures human knowledge of AI and experience in designing and using

AI. This knowledge scale specifically evaluated students' understanding of the inputs and outputs of generative AI, focusing on their comprehension of prompts (e.g.,"*I have knowledge of the prompt requirements for generative AI*") and their relationship with generative AI's outputs (e.g., "*I have knowledge of which generative AI outputs are attainable with current methods*"). The instrument employed a seven-point Likert scale and demonstrated high reliability, with a Cronbach's alpha value of 0.936.

### *3.3.2. Qualitative Data*

Qualitative data to explore the workshop's impact on prompt engineering skills comprised pre- and post-workshop prompt libraries. Each prompt library was a sheet on a Google spreadsheet editable by participants. Each sheet comprised eight column headings: Author; [Prompt] Title; Prompt; Difficulty Level; Best Chatbot; Worst Chatbot; Link to Best Input and Output; and Link to Worst Input and Output. The author column was pre-filled with participants' names.

At the beginning of the workshop, participants were given five minutes to attempt the task found in the guided practice module, that is, to engineer prompts to plan the course's final essay task (see Appendix B). Participants attempted the task in their respective rows in the pre-workshop prompt library sheet, by either writing their prompts under the Prompt column heading or pasting links to their chatbot conversations. If participants tested prompts on chatbots, they were encouraged to name the best and worst chatbots. Participants could use whatever chatbot or chatbot software they preferred. Immediately before taking the post-workshop questionnaire, participants attempted the task again (See Appendix C).

Qualitative data to provide further insight into AI self-efficacy and knowledge of generative AI comprised end-of-term reflections. The reflections were a compulsory assignment that students submitted at the end of the three-week course. The assignment was open-ended (see Appendix D) so students could write about the workshop but had no obligation to do so.

### 3.4. Data Analysis
### *3.4.1. Quantitative Data*
To evaluate the effects of the prompt engineering workshop on students' AI self-efficacy (RQ1) and generative AI knowledge (RQ2), we employed the Wilcoxon signed-rank test to determine whether there were significant differences in these constructs before and after the workshop. Although 27 students attended the workshop, only 19 completed both the pre- and post-workshop questionnaires. Given the limited sample size, the non-parametric significance test was used to assess changes in students' AI self-efficacy and generative AI knowledge.

### *3.4.2. Qualitative Data*
To evaluate the workshop's influence on students' ability to engineer prompts (RQ3), we analyzed the pre- and post-workshop prompt libraries. First, we used a directed approach (Hsieh & Shannon, 2005). We designed an initial coding scheme of the strategies we had introduced in the guided practice module. These include core strategies we had highlighted in the guided practice module and optional strategies that we challenged students to apply. Each student's prompt library's content was then coded for the presence or absence of each core and optional strategy. To enhance reliability, the first and third authors independently coded the libraries. They discussed and resolved any coding discrepancies and wrote a codebook.

Second, we used a summative approach to evaluate the sophistication of each student's prompting (Hsieh & Shannon, 2005). We counted the number of strategies in a student's pre- and post-workshop prompt libraries and compared the results. Furthermore, we compiled descriptive statistics to compare each strategy's usage before and after the workshop and all students' performance before and after the workshop.

For insights into AI self-efficacy and generative AI knowledge from the end-of-term reflections, the first and third authors performed a conventional content analysis. They read students' reflections and identified three that mentioned the workshop. They inductively categorized each as addressing either RQ1, RQ2 or both, and then specified in what way.

## 4. Results

### 4.1. Effects on AI Self-Efficacy (RQ1)

Table 3 presents the results of the Wilcoxon signed-rank test comparing students' AI self-efficacy before and after the workshop. Before the workshop, students reported a mean score of 4.600 (SD = 1.021), whereas after the workshop, they reported a mean score of 5.084 (SD = 0.963). The Wilcoxon signed-rank test result (z = 1.784, p > 0.05) indicates no significant difference in students' AI self-efficacy before and after the workshop. However, the observed increase in both the mean and median scores of students' AI self-efficacy after the workshop suggests a partial improvement, indicating the potential effectiveness of the prompt engineering workshop.

Similarly, Wilcoxon signed-rank tests were conducted on the two specific dimensions of comfort with generative AI and technological skills. While no significant differences were found in these two dimensions, it is worth noting that both the mean and median scores for comfort with generative AI and technological skills improved after the workshop. This improvement is further demonstrated by the increased mean scores for all items in Figure 1 and 2, reflecting an overall positive trend in students' perceptions of their AI self-efficacy.

**Table 3.** The Wilcoxon signed-rank test comparing students' AI self-efficacy before and after the workshop.

|  |  | N | Median | Mean (SD) | Z | Significance |
|---|---|---|---|---|---|---|
| AI self-efficacy | Pre-workshop | 19 | 4.600 | 4.716 (1.021) | 1.784 | 0.074 |
|  | Post-workshop | 19 | 5.000 | 5.084 (0.963) |  |  |
| Comfort with GAI | Pre-workshop | 19 | 4.667 | 4.737 (1.147) | 1.478 | 0.139 |
|  | Post-workshop | 19 | 5.000 | 5.070 (1.081) |  |  |
| Technological skills | Pre-workshop | 19 | 4.500 | 4.684 (1.118) | 1.511 | 0.131 |
|  | Post-workshop | 19 | 5.000 | 5.105 (0.959) |  |  |

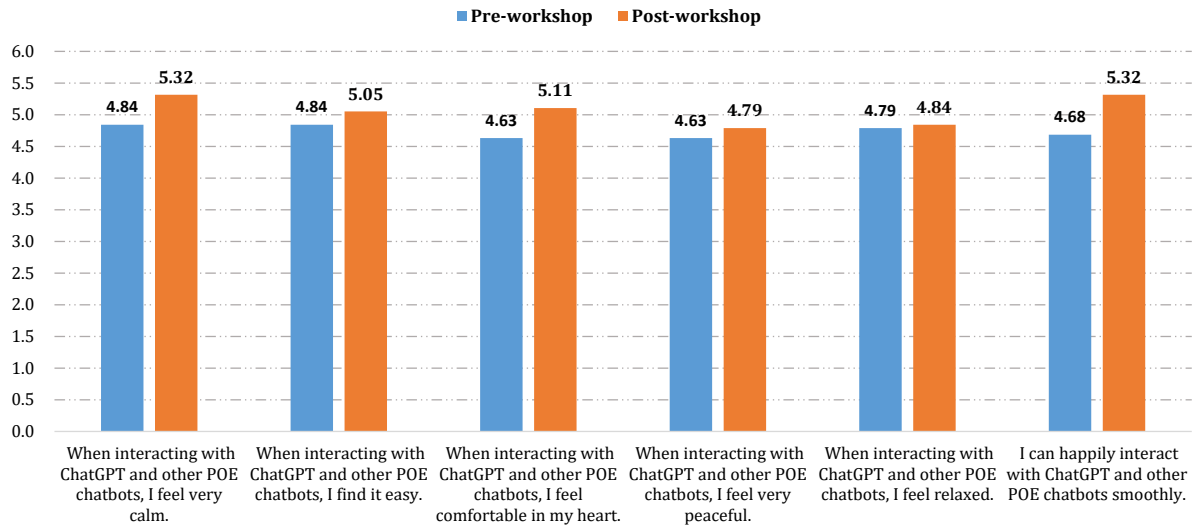

**Figure 1.** The mean score on all items of comfort with generative AI before and after the workshop.

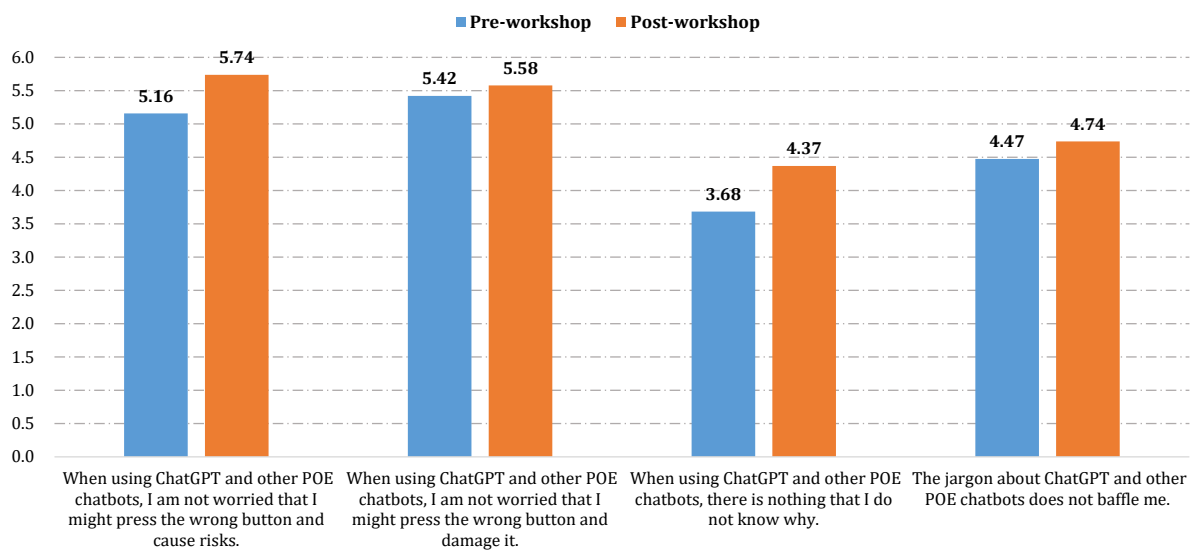

**Figure 2.** The mean score on all items of technological skills before and after the workshop.

We found two end-of-term reflections that cited the first author's detailed instructions for the workshop's impact on AI self-efficacy. For Paul, this instruction was beneficial for his AI self-efficacy:

> *As [the first author] guided us down each step of prompting AI for constructive responses, I now feel less stressed and more confident in writing a final essay of high quality which I have always found myself lacked proficiency in. I am now excited to try blending AI into different learning tasks...*

On the other hand, for Belle, a science student who had not used AI much, the detailed instructions raised ethical concerns that appeared not so beneficial to AI self-efficacy:

> *[The first author] gave such a detailed course on how to make ChatGPT perform better to assist on the final essay. But I kept thinking that if I use the such detailed information and steps, ignoring the fact that I am the one writing the words and phrases, the whole soul and ideal of* *the essay is I don't see humanity in the essay, or maybe its humanity comes from the creator of the AI, but that's not me. Even despite the fact of arguing whether it is written by the AI or by me following the AI's order, won't it be an academic misconduct? The questionaire asked whether or not I feel relaxed when using ChatGPTs, I just felt tense and stress.*

### 4.2. Effects on AI Knowledge (RQ2)

Table 4 presents the results of the Wilcoxon signed-rank test, which compares students' generative AI knowledge before and after the workshop. Prior to the workshop, students exhibited a mean score of 3.825 (SD = 1.364), indicating a low level of understanding regarding prompts and outputs in generative AI. In contrast, after the workshop, they reported a mean score of 5.281 (SD = 0.824). The Wilcoxon signed-rank test result (z = 3.377, p < 0.001) indicates a significant increase in students' perceived generative AI knowledge following the workshop. Figure 3 visually depicts the change in scores for each item, with all items demonstrating a notable increase after the workshop. These findings suggest that the prompt engineering workshop effectively enhanced students' generative AI knowledge.

**Table 4.** The Wilcoxon signed-rank test comparing students' generative AI knowledge before and after the workshop.

|  | N | Median | Mean (SD) | Z | Significance |
|---|---|---|---|---|---|
| Pre-workshop | 19 | 3.667 | 3.825 (1.364) | 3.377 | < 0.001 |
| Post-workshop | 19 | 5.167 | 5.281 (0.824) |  |  |

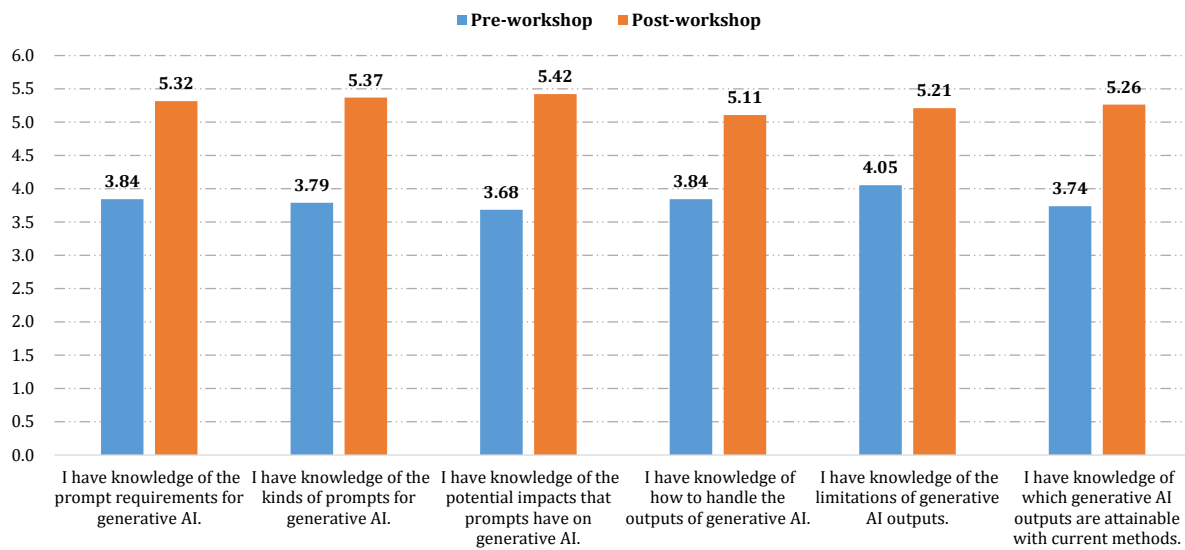

**Figure 3.** The mean score on all items of generative AI knowledge before and after the workshop.

We found one end-of-term reflection that described the workshop's influence on knowledge of generative AI. Apple, a social sciences student, cited applying a prompt engineering strategy to the use case had benefited her knowledge of generative AI:

*The ChatGPT prompt engineering workshop is very practical, and I left the classroom with a lot of skills that I could put into practice into future. I am kind of a heavy AI tool user, especially it helped my work in some aspects. However, there are difficulty on how to properly prompt the ChatGPT, and I tried spent a lot of time yet still cannot get what I need from the tool. After this*

*session I've learnt to prompt with workflow/steps so it could perform as I'd like it to. Learning to utilize AI tools is very crucial and it can definitely become an edge for us. The part when we are guided to ask ChatGPT for analysis the difficulty of different essay questions is really thoughtful. I never thought of such usage, and it could help us to pick the right topic to work on based on our strengths and sources available. This is a really helpful session and I think all students in HKU should attend this workshop!*

### 4.3. Effects on Prompt Engineering Ability (RQ3)

Appendix E lists the 13 prompt engineering strategies introduced in the workshop's guided practice module and categorized as either a core or an optional strategy. Each strategy is listed with its description, example prompts taken from the worksheets, and the number of students who employed the strategy in the pre- and post-workshop libraries. At the bottom of Appendix E, we observe that the total number of strategy instances found in the pre-workshop library was 18 and that number increased in the post-workshop library to 52. 24 students attempted the pre-workshop prompt library and 14 the post-workshop by inputting any information into their rows. When considering the number of participating students in each library, we find the average number of strategies used by a student was 0.75 in the pre-workshop library but 3.71 in the post-workshop library.

13 students attempted both the pre- and post-workshop libraries. One student submitted a broken link in the post-workshop library that could not be analyzed. Two students (Mimi and Nora) reduced their strategy instances from one in the pre-workshop to zero in the post-workshop. Another two students, Gina and Melvin, maintained the same number of strategy instances in the pre- and post-workshop libraries, at zero and two, respectively. Additionally, Melvin had used different strategy combinations for the pre-workshop library (code nos. 4 and 13) and the post-workshop (code nos. 1 and 4). Importantly, we found seven students had increased their pre- to post-workshop strategy instances. Some students showed remarkable gains. For instance, Joyce and Paul showed zero strategies in their pre-workshop libraries but 11 in their post-workshop libraries. Similarly, Gail and Zeke had shown one strategy in their pre-workshop libraries but nine and 10, respectively, in their post-workshop libraries. These four students came from Humanities, Science, Social Sciences, and Science backgrounds respectively, which may suggest that students could develop prompt engineering competency irrespective of their study background.

In the pre-workshop library, we observed students had written prompts directly in the sheet. In the post-workshop library, students pasted links to chatbot conversations from poe.com, chatgpt.com or kimi.moonshot.cn. In those conversations, we observed that students had demonstrated strategies by copying prompt templates from the worksheets and pasting them verbatim into a chatbot. For example, in Figure 4, Gail had pasted verbatim a worksheet prompt into Mixtral-8x7B-Chat.

**Figure 4**. Gail's conversations with Chatbot.

We also observed students employed strategies by slightly modifying prompt templates from the worksheets, for instance, by selecting an academic source for the chatbot to summarize. In Figure 5, Edna modified a worksheet prompt by inputting her chosen final essay question instead of that given in the worksheets.

I have chosen the final essay question: Question 4: 'Hong Kong in essence remains what it has always been, a market place.' (Endacott, 1958) Do you agree with this depiction of Hong Kong history?

Generate an outline for the final essay question. Before you generate the outline, first, think of
the key events, figures, or concepts I should research to answer my question. Then organize that
information coherently (e.g. chronologically) into sections of an essay (e.g. introduction, main
arguments, conclusion).

The outline must include:

1. a potential thesis statement
2. three or four main arguments
3. suggestions for evidence to support each argument
4. steps to complete each section
5. suggested word count for each section

Generate the outline in a table format. The table should comprise three columns with the
headings:

Talk to GPT-3.5-Turbo on Poe

→| Continue chat        ✎ New chat

**Figure 5**. Edna's conversations with Chatbot.

Finally, we observed students who employed strategies by writing prompts not given in the worksheets. For example, in <span style="color:blue">Figure 6</span>, Paul had instructed GPT-4o-128k to rewrite an essay outline by adding page numbers to academic sources.

**Figure 6**. Paul's conversations with Chatbot.

# 5. Discussion

## 5.1. Major Findings

The findings from 27 undergraduate students reveal substantial changes in students' AI self-efficacy, knowledge of generative AI and abilities to engineer effective prompts after a prompt engineering intervention. This appears irrespective of their major or field of study, echoing findings from Kong et al. (2021) that showed students from various disciplines could benefit from an AI literacy course. The findings provide valuable insights that can inform the design and refinement of prompt engineering education.

Students reported a higher level of AI self-efficacy after the workshop than before, although the difference was not statistically significant. Thus, interacting with chatbots and completing the required tasks somewhat improved students' comfort with AI as well as technological skills. Two factors could explain that. On one hand, the nature of generative AI chatbots generally provided students with a smooth experience, fostering feelings of comfort and confidence. For instance, as Ouyang et al. (2020) indicated, current LLMs aim to produce human-like, helpful, and safe content while avoiding harmful outputs. The anthropomorphic and conversational features of chatbots are proven to provide a smooth user experience, enabling users to trust and accept these AI technologies (Glikson & Woolley, 2020; D. Wang et al., 2024). Despite this, some students, like the science student named Belle, reported feeling tense and stressed when using generative AI for academic writing due to concerns about ethical issues such as misconduct, which potentially undermined their comfort with AI. Therefore, addressing these ethical concerns should be a priority in future prompt engineering education to significantly increase students' AI self-efficacy. On the other hand, the improvement in students' comfort with AI and technological skills can also be attributed to the effective design of the instructional activities in the workshop. The detailed instructional procedures and appropriately arranged learning tasks did not overwhelm the students but rather boosted their confidence, as evidenced by reflections from the student named Paul. Ultimately, since AI self-efficacy has been shown to impact students' acceptance and intention to use AI (Chai et al., 2021; Kwak et al., 2022; Chen et al., 2024), an implication of students' enhanced AI self-efficacy from the prompt engineering workshop is that students might continue not only using AI but also engineering prompts. That puts students in an advantageous position to continue gaining skills to navigate the evolving landscape of AI applications.

Second, students demonstrated a significantly higher level of generative AI knowledge after the workshop than before. As shown in Figure 3, they expressed that after attending the workshop, they better understood the elements, types, and strategies of prompts and the relationship between prompts and generative AI outputs. In the post-workshop reflection stage, students like Apple, a social sciences student, expressed that before the workshop, they faced considerable difficulties in generating appropriate prompts for ChatGPT. However, after completing the workshop, they acquired knowledge to engineer effective prompts tailored to their needs. These results suggest that the designed prompt engineering workshop significantly improved students' generative AI knowledge, particularly in crafting prompts, aligning with Knoth et al.'s (2024) findings that prompt engineering knowledge and skills can be effectively acquired through instruction and training.

Third, beyond self-reported data, students' ability to engineer proper prompts was also found to have improved in practice. The total number of effective strategy instances increased from 18 in the pre-workshop library to 52 in the post-workshop library. The average number of strategies used per student rose from 0.75 to 3.71, indicating that students had engineered more sophisticated prompts. Like the questionnaire and self-reflection results, the prompt libraries indicate that prompt engineering is a trainable skill for students in higher education. Besides, our analysis of prompt libraries contributes to mixed methods approaches to understanding actual ChatGPT usage in higher education contexts (Baig & Yadegaridehkordi, 2024).

In the pre-workshop library, we observed that the strategy of reference texts was the most popular followed by the persona strategy. The popularity of the former may come from the students' perceived need to use external information, such as the final essay questions, in their prompts; the latter may come from previous exposure to the strategy (Prompt engineering, n.d.). In the post-workshop library, we observed that the total number of each strategy instance closely followed the scaffolding of the strategies in the guided practice module. This suggests that students followed the instructor step-by-step to apply strategies. That aligns with the end-of-term reflections which suggest students appreciated this step-by-step approach.

Four students showed extraordinary positive changes in their prompt engineering skills. This suggests that the workshop design and implementation were very effective for these students. On the other hand, most students did not show such drastic positive change. Besides, many students did not contribute to the post-workshop when they had to build the pre-workshop library. Therefore, although students self-reported increases in AI self-efficacy, the workshop instruction could be improved so more students would demonstrate their knowledge and skills.

In sum, deductive coding of prompt libraries has shown that prompt engineering is a trainable skill for students in higher education. Our study's workshop has contributed to higher education efforts (Eager & Brunton, 2023; Lo, 2023) to provide students with structured prompt engineering guidance successfully. Because of these efforts, students are moving from intuitive, ad hoc learning (Zamfirescu-Pereira et al., 2023) to explicitly and systematically using effective and sophisticated prompt engineering strategies. These results also resonate with the students' self-reported increases in AI self-efficacy and their end-of-term reflections. Like an extensive AI literacy course for undergraduate students (Kong et al., 2022), our study demonstrates that an abbreviated, focused and detailed prompt engineering course can enhance students' AI self-efficacy and facilitate their rapid adoption of effective prompt engineering strategies.

## 5.2. Limitations and Future Research

The relationship between AI self-efficacy, generative AI knowledge and prompt engineering ability should be further investigated as it has implications for approaching AI education in higher education. First, the sample size was small and although its participants came from various disciplines, their use case was narrow, confined to a history course's final essay task. To increase the generalization of findings, prompt engineering interventions can be designed, implemented and evaluated with more students and in different higher education disciplines. Second, this study's prompt engineering intervention was a one-off, and there has been no follow-up on the sustainability of students' AI self-efficacy gains and prompt engineering skills. Future research could explore whether these students apply their skills to other coursework in higher education. Third, the intervention was isolated in its university context. The intervention could be integrated into broader prompt engineering education that aligns with a higher education institution's overall AI curriculum design and development.

## 6. Conclusion

With the advancement of generative AI, prompt engineering has become an increasingly important skill that can enhance students' learning and daily life. To equip students with this skill, this study designed and evaluated a prompt engineering intervention at a university in Hong Kong. The results show that a prompt engineering workshop can substantially impact undergraduate students' AI self-efficacy and generative AI knowledge. It can enhance these students' ability to engineer effective prompts for their academic writing tasks. The self-reported improvements in AI self-efficacy and generative AI knowledge and the observed improvements in effective prompt engineering suggest that a brief, targeted intervention can rapidly equip students with AI literacy for their context-specific academic writing.

**Appendix A**
The course's final essay task

Final Essay
Your final essay should be 1,500-2,000 words (excluding footnotes), with proper footnotes and bibliography. You are expected to use at least SIX published academic sources (websites, news articles, and lecture slides do not fall in this category).

You should indicate your full name, university number, and use your selected question in full as the title. Your final essay should be uploaded to Moodle by 9pm on 19 July.

1. Has Hong Kong's economic history always been tied to the mainland?
2. Who or what were the most important sources of authority in Hong Kong government between 1842 and 1997?
3. Why has it been so difficult to implement comprehensive social welfare throughout Hong Kong history?
4. 'Hong Kong in essence remains what it has always been, a market place.' (Endacott, 1958) Do you agree with this depiction of Hong Kong history?
5. When and why did the British Hong Kong government become more invested in the people of Hong Kong (if at all)?
6. To what extent did 'Hong Kong identity' become more pronounced only after the Sino-British Joint Declaration?
7. What makes presenting Hong Kong history so difficult? (Answers must address academic texts but may also discuss non-academic representations.)
8. 'From Fishing Village to Modern Metropolis.' Do you agree with this depiction of Hong Kong's historical trajectory?

**Appendix B**
Pre-workshop prompt exercise

Write a prompt or prompts for ChatGPT to plan your HIST1017 final essay task.

- In 2_HIST107S Prompt Library, submit the prompt(s) to the pre-workshop tab, column C
- If you test the prompt(s) on chatbots, kindly complete columns E, F, G and H

**Appendix C**
Post-workshop prompt exercise

Write a prompt or prompts for ChatGPT to plan your HIST1017 final essay task.

- In 2_HIST107S Prompt Library, please paste the link to your prompts and chatbot output to the post-workshop tab, column G
- If you test the prompt(s) on chatbots, kindly complete columns E, F and H

**Appendix D**
The open-ended course assignment

The cumulative journal is for students to note down their learning from each day's learning activities and readings. Students need not write all the things they have learned, but should share what things they found interesting, especially that which challenged their previous understanding of Hong Kong history. Students are strongly advised to write in their journal regularly throughout the course. The total wordcount of the journal should be no more than 1,200 words and should be uploaded to Moodle by 9pm on 13 July. Footnotes are optional.

Prompt engineering strategies taught by the instructor and used by the students

| No. | Type | Code | Description | Example Prompt(s) | Pre-workshop count | Post-workshop count |
|---|---|---|---|---|---|---|
| 1 | Core | Chain prompt | Uses more than one prompt to break down a complex task into sequential, smaller tasks or long documents into smaller chunks | ###Prompt 1<br>Your task is to help me plan my history final essay.<br><br>Here are my history final essay questions:<br><br><questions><br>[insert questions]<br></questions><br><br>Which question would be the most interesting and feasible for a student without a strong background in history to answer? When you provide an answer, please explain the reasoning and assumptions behind your answer.<br><br>###Prompt 2<br>I have chosen the final essay question: [insert chosen question]. question 5: "When and why did the British Hong Kong government become more invested in the people of Hong Kong (if at all)?"<br><br>Generate an outline for the final essay question. Before you generate the outline, first, think of the key events, figures, or concepts I should research to answer my question. Then organize that information coherently (e.g. chronologically) into sections of an essay (e.g. introduction, main arguments, conclusion). | 0 | 5 |
| 2 | Core | Delimiters | Uses visualizers to help the chatbot to understand distinct parts of input (e.g. XML tags such as <document></document> and <example></example>; Markdown syntax such as # before a first heading, ## before a second heading; and punctuation marks) | Here are my history final essay questions:<br><br><questions><br>[insert questions]<br></questions> | 0 | 5 |
| 3 | Core | External tools | Creates a custom chatbot with a system prompt and knowledge base [to automate a prompt engineering process] | N/A | 0 | 2 |

| 4 | Core | Reference texts | Attaches a file or pastes external text verbatim that the chatbout should use as input (i.e. for retrieval augmented generation) | ###Prompt 1<br>Your task is to help me plan my history final essay.<br><br>Here are my history final essay questions:<br><br><questions><br>[insert questions]<br></questions><br><br>Which question would be the most interesting and feasible for a student without a strong background in history to answer? When you provide an answer, please explain the reasoning and assumptions behind your answer.<br><br>###Prompt 2<br>To address any potential ambiguities or limitations in your answer, I have attached my history course outline. Furthermore, ask me at most four questions that would help you produce a better answer. After you receive my answers, cite examples or evidence from my answers and the course outline to suggest a better version of your answer.<br><br>[attach course outline]<br><br>###Prompt 3<br>Your task is to evaluate three examples of history final essays: Essay 1; Essay 2; and Essay 3.<br><br>#Evaluation rules<br>[insert evaluation details]<br>Each essay includes the main body text, bibliography, grade and rubric descriptor for that grade. Essay 1 and Essay 2 also include a word count.<br><br>Identify three to five features found in Essay 1 and Essay 2, but not in | 12 | 7 |

Essay 3. The features may be related to the content, language or organization of the essay. If possible, use specific examples or evidence from the essays to support your explanation of the features.

The essays, the main body texts, the bibliographies, the word counts, the grades and the rubric descriptors are delimited by XML tags below:

\<insert each final essay, delimited by \<essay>\</essay>>
\<insert main body text, delimited by \<text>\</text>>
\<insert word count, delimited by \<word count>\</word count>>
\<insert bibliography, delimited by \\>
\<insert grade, delimited by \<grade>\</grade>>
\<insert rubric descriptor, delimited by \<rubric descriptor>\</rubric descriptor>>

| 5 | Core | Systematically test | Tests prompt(s) on more than one chatbot, evaluating chatbots' output. Recommends best and worst chatbots in prompt library or links to best and worst input and output. | N/A | 0 | 0 |
|---|---|---|---|---|---|---|
| 6 | Optional | Chain of thought prompting | Provide intermediate reasoning steps in examples so that the chatbot acquires intermediate reasoning ability; or prompts the chatbot to think step-by-step [before generating output] | ###Prompt 1<br>[insert instructions on how to revise the outline]<br>Your next task is to rewrite the essay outline, incorporating information from the six academic sources into the outline. Before you rewrite the outline, first consider which section(s) of the essay each academic source can support and how.<br><br>###Prompt 2<br>After you receive my answers, but before drafting the section, explain your reasoning and assumptions step-by-step in tags. | 0 | 5 |
| 7 | Optional | Persona | Instructs ChatGPT to play a customized persona or role when generating output | ###Prompt 1<br>You are a helpful Hong Kong history instructor and writing consultant for the Department of History at the University of Hong Kong.<br><br>###Prompt 2<br>From now on, act as a...Provide outputs that a...would...<br><br>###Prompt 3<br>You are going to pretend to be a...you are going to output the corresponding text that...would produce. | 4 | 4 |

| 8 | Optional | Question refinement | Instructs ChatGPT to improve the quality of the input and output by its suggesting a better version of [something] | ###Prompt 1<br>From now on, whenever I ask a question, ask four additional questions that would help you produce a better version of my original question. Then, use my answers to suggest a better version of my original question.<br><br>###Prompt 2<br>To address any potential ambiguities or limitations in your answer, I have attached my history course outline. Furthermore, ask me at most four questions that would help you produce a better answer. After you receive my answers, cite examples or evidence from my answers and the course outline to suggest a better version of your answer.<br><br>[attach course outline] | 0 | 4 |
|---|---|---|---|---|---|---|
| 9 | Optional | Rewrite | Instructs ChatGPT to rewrite a text, for instance, by adding or translating information | ###Prompt 1<br>Rewrite the following text to be more serious:<br><br>—<br><br>{very informal text}<br><br>—<br><br>###Prompt 2<br>[insert instructions on how to revise the outline]<br>Your next task is to rewrite the essay outline, incorporating information from the six academic sources into the outline. Before you rewrite the outline, first consider which section(s) of the essay each academic source can support and how. | 0 | 4 |

| 10 | Optional | Reflection | Instructs ChatGPT to introspect on its output and identify any errors | ###Prompt 1<br>When you provide an answer, please explain the reasoning and assumptions behind your answer. If possible, use specific examples or evidence to support your explanation of why the answer is the best. Moreover, please address any potential ambiguities or limitations in your answer, in order to provide a more complete and accurate response.<br><br>###Prompt 2<br>Which question would be the most interesting and feasible for a student without a strong background in history to answer? When you provide an answer, please explain the reasoning and assumptions behind your answer. | 0 | 4 |
|---|---|---|---|---|---|---|
| 11 | Optional | Summarization | Instructs ChatGPT to summarize a text | ###Prompt 1<br>{chat transcript}<br><br>Summarize the above conversation between a teacher and student. Make sure to state any learning difficulties that the student has.<br><br>###Prompt 2<br>Your task is to summarize the attached academic source: [name academic source]<br>Chapters 4 and 5 of Steve Tsang's A Modern History of Hong Kong (2004).<br><br>#Summary rules<br>[insert details about the summary's content and structure]<br>Summarize the academic source concisely and include relevant details for the following:<br>1.      The initial British approach to governing Hong Kong<br>2.      Key reforms pre-World War 2 (WWII) | 0 | 3 |

| 12 | Optional | Template | Instructs ChatGPT to fill in a user-specified template with content | ###Prompt 1<br>I am going to provide a template for your output. Everything in all caps is a placeholder. Any time that you generate text, try to fit it into one of the placeholders that I list. Please preserve the formatting and overall template that I provide.<br><br>###Prompt 2<br>#Template rules<br>[prefill output or provide details about output format]<br>Output your summary as bullet points that I can easily paste into my essay. Output the bullet points under the following headings:<br><br>The initial British approach to governing Hong Kong:<br><br>Key reforms pre-World War 2 (WWII): | 1 | 3 |

| 13 | Optional | Visualization generator | Instructs ChatGPT to visualize textual outputs | #Prompt 1<br>[insert details about visualizing the output]<br>Generate the outline in a table format. The table should comprise three columns with the headings: 1. Section<br>2. Description<br>3. Steps<br>In the Section column, name each section. In the Description column, state the purpose of the section, its key points, supporting evidence and target word count. In the Steps column, provide a sequence of at most five steps in which I might complete the section. Generate the table outline without XML tags:<br><br>###Prompt 2<br>[insert details about visualizing the output]<br>To visualize the revised outline, add a fourth column with the heading: academic sources. In this column, list the relevant academic sources for each section and what I should read in each source.<br><br>###Prompt 3<br>#Template rules<br>[prefill output or provide details about output format]<br>Output your summary as bullet points that I can easily paste into my essay. Output the bullet points under the following headings:<br><br>The initial British approach to governing Hong Kong:<br><br>Key reforms pre-World War 2 (WWII): | 1 | 6 |
| | | | Total No. of Strategy Instances | | 18 | 52 |
| | | | Mean No. of Strategy Instances per Participating Student | | 0.75 | 3.71 |